\documentclass{appolb}
\usepackage{epsfig}
\usepackage[utf8]{inputenc}
\usepackage[OT4]{fontenc}
\usepackage[MeX]{polski}
\usepackage[polish,english]{babel}
\usepackage{xcolor}

\definecolor{olive}{rgb}{0.3, 0.4, .1}
\definecolor{fore}{RGB}{249,242,215}
\definecolor{back}{RGB}{51,51,51}
\definecolor{title}{RGB}{255,0,90}
\definecolor{dgreen}{rgb}{0.,0.6,0.}
\definecolor{gold}{rgb}{1.,0.84,0.}
\definecolor{JungleGreen}{cmyk}{0.99,0,0.52,0}
\definecolor{BlueGreen}{cmyk}{0.85,0,0.33,0}
\definecolor{RawSienna}{cmyk}{0,0.72,1,0.45}
\definecolor{Magenta}{cmyk}{0,1,0,0}

\newcommand{\twidth}{0.32\textwidth}%

\begin{document}


\title{Pion spectra in Ar+Sc interactions at SPS energies%
\thanks{Presented at the Critical Point and Onset of Deconfinement 2016, Wroclaw, Poland,
May 30th - June 4th, 2016}%
}
\author{Maciej Lewicki for the NA61/SHINE Collaboration
\address{Faculty of Physics and Astronomy, University of Wrocław, pl. Uniwersytecki 1, 50-137 Wrocław, Poland}
}

\maketitle
\begin{abstract}
This contribution discusses recent results from analysis of Ar+Sc interactions recorded with the NA61/SHINE detector at six beam momenta: 13A, 19A, 30A, 40A, 75A, 150A GeV/c at the CERN SPS. Rapidity and transverse mass spectra of pions obtained with the "h-" analysis method are presented and compared to results from p+p, Be+Be and Pb+Pb collisions.
\end{abstract}

\section{Introduction}
NA61/SHINE is a fixed target spectrometer located in CERN's North Area utilizing the SPS proton, ion and hadron beams \cite{facility}. Tracking capabilities are provided by four large volume Time Projection Chambers (TPC), two of which are located in the magnetic field. The Projectile Spectator Detector (PSD) -- a zero degree, modular calorimeter, is used to determine centrality of the collisions.\\
\indent The aim of the experiment is to explore the QCD phase diagram $(\mu_B,T)$ by a two-dimensional scan in collision energy and system size. The yields of hadrons produced in the collisions are studied for indications of the onset of deconfinement (\textit{kink}, \textit{horn} and \textit{step}) and the critical point of the phase transition (hill in event-by-event fluctuations). Recent measurements of Argon and Scandium collisions are an important step in this program. Comparison of negative pion spectra give important insight into the dynamics of ion collisions at the border of light and heavy nuclei.

\section{h$^-$ Method}
Negatively charged pion spectra were measured using the so called "h$^-$ method". It exploits the fact that $\approx90\%$ of negative hadrons produced in the SPS energy range are $\pi^-$ mesons. The small ($\approx10\%$) contribution of other particles ($K^-, \bar{p}$) was subtracted from negative pion spectra using simulations with the EPOS-1.99 model~\cite{EPOS}. The latter was tested on proton-proton data and showed only small discrepancies in the ratio $K^-/\pi^-$ ($K^-$ are the main contribution to the correction)\cite{antoni}.

\section{Data analysis}
The events recorded by the NA61/SHINE spectrometer were selected for data quality and centrality of the collisions. Centrality classes were determined using the PSD, located most downstream on the beam line It measures predominantly the projectile spectator forward energy $E_F$ of the non-interacting nucleons of the beam nucleus. The distibution of $E_F$ was used to define and select event classes corresponding to the collision centrality.

Corrections of the raw data was based on EPOS-1.99~\cite{EPOS} (version CRMC 1.5.3.) model simulation of hadron interactions and the GEANT-3.2 code for particle transport and detector simulation (see~\cite{pp}). Centrality classes in the model calculations were selected by the number of forward spectator nucleons.\\
\indent Results presented in the plots are shown only with statistical uncerainties. These come from two sources: the experimental data and the simulation based corrections. The contribution of the latter is insignificant ($<0.1$\%).
Based on the previous analysis of Be+Be\cite{BeBe} and p+p\cite{pp} reactions, we estimate systematic errors at a level of 5\%-10\%.
\section{Rapidity and Transverse Momentum Spectra}
\begin{figure}[h]
\includegraphics[width=\twidth]{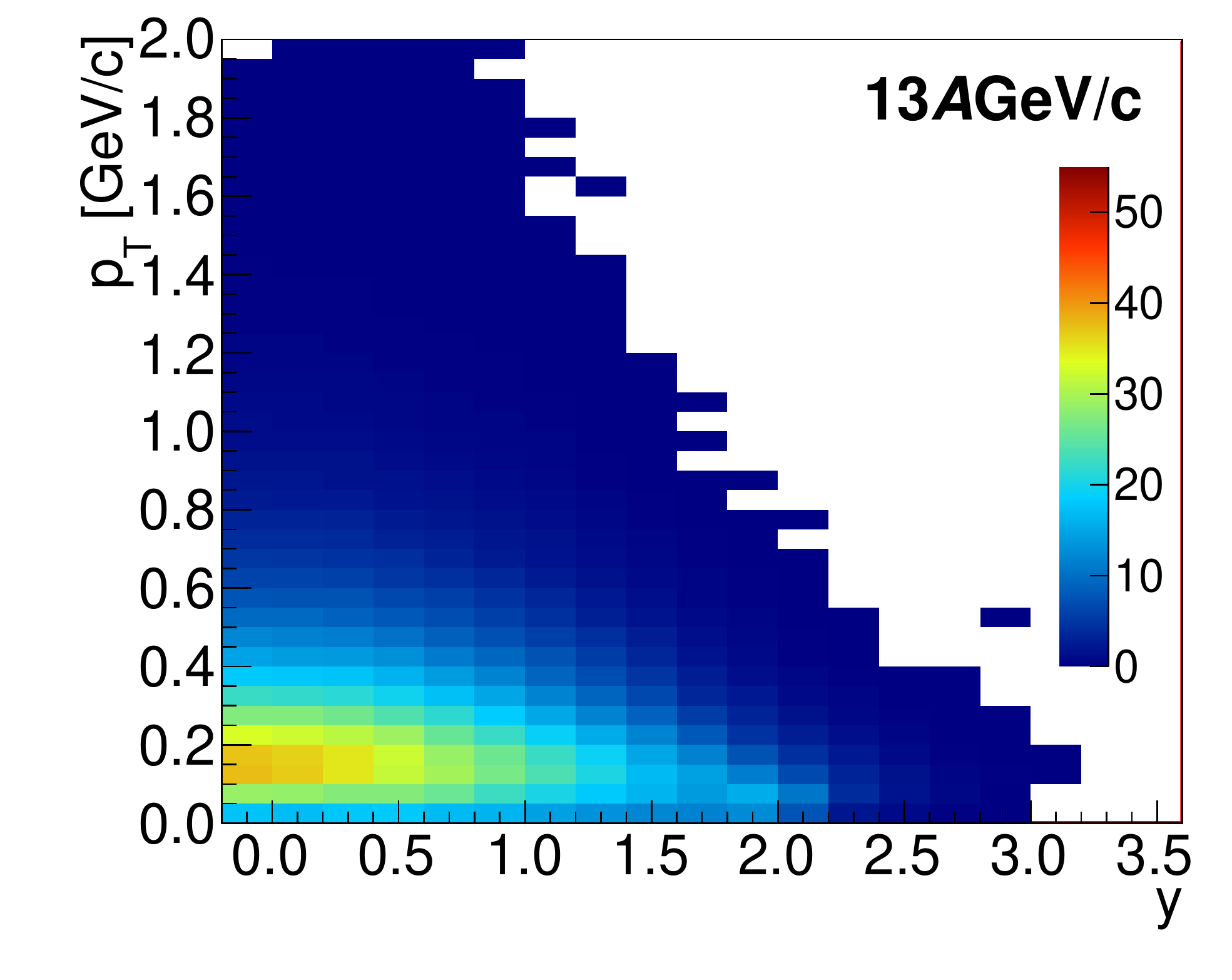}
\includegraphics[width=\twidth]{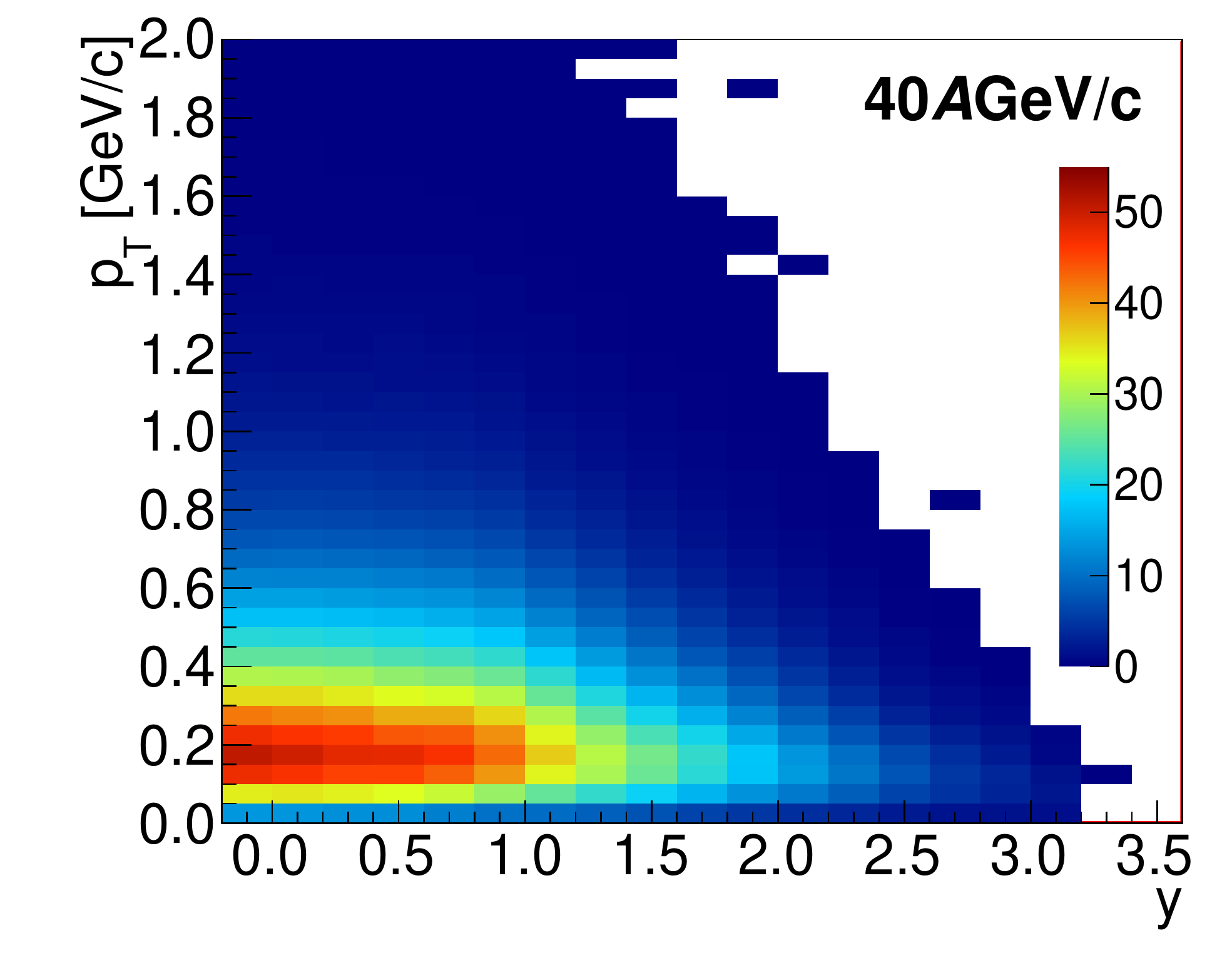}
\includegraphics[width=\twidth]{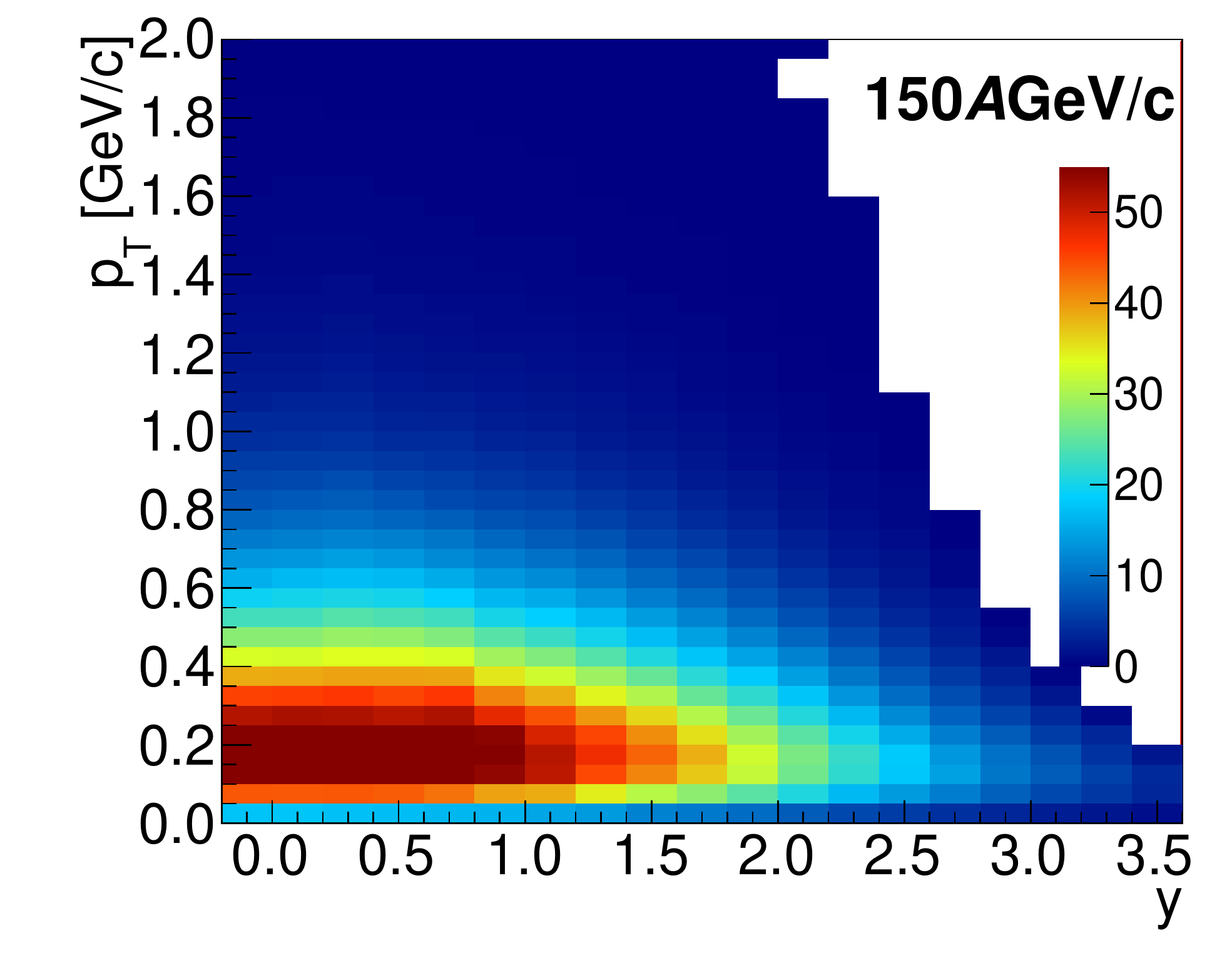}
\caption{\textit{Measurements of the double differential spectra $\left(\frac{dn^2}{dy~dp_T}\right)$ of negatively charged pions in central Ar+Sc collisions.}}
\label{fig:dd_spectra}
\end{figure}

Results, corrected for acceptance and background, were obtained for the 5\% most central Ar+Sc interactions at beam energies of $19A$, $30A$, $40A$, $75A$ and $150A$ GeV/c. These are compared to measurements from inelastic p+p~\cite{pp} and the 5\% most central Be+Be collisions~\cite{BeBe} from NA61/SHINE as well as from central (5 \%or 7\%) Pb+Pb collisions from NA49~\cite{PbPbHigh, PbPbLow}. 
\indent Preliminary results on double differential negatively charged pion spectra are presented in Fig.\ref{fig:dd_spectra} for beam energies of $19A$, $40A$, and $150A$ GeV/c. They cover the full forward rapidity hemisphere.
\indent In order to obtain rapidity distributions $\frac{dn}{dy}$, the data was extrapolated beyond the detector accptance. The extrapolation in the direction of $p_T$ from the acceptance edge to the $p_T=3.0$~GeV/c was performed with an exponential dependence fitted in the measured region.\\
\begin{figure}[h]
\includegraphics[width=\twidth]{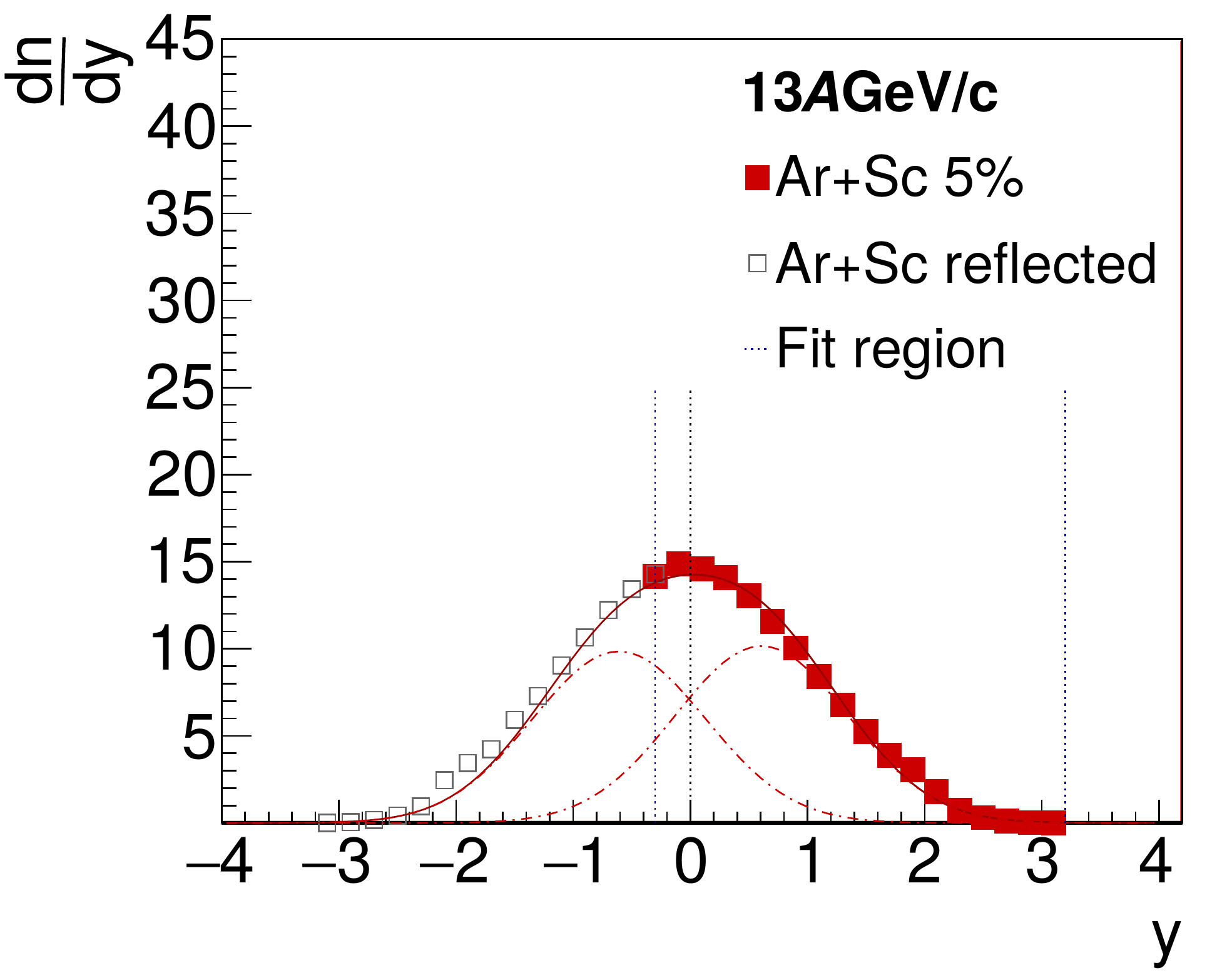} 
\includegraphics[width=\twidth]{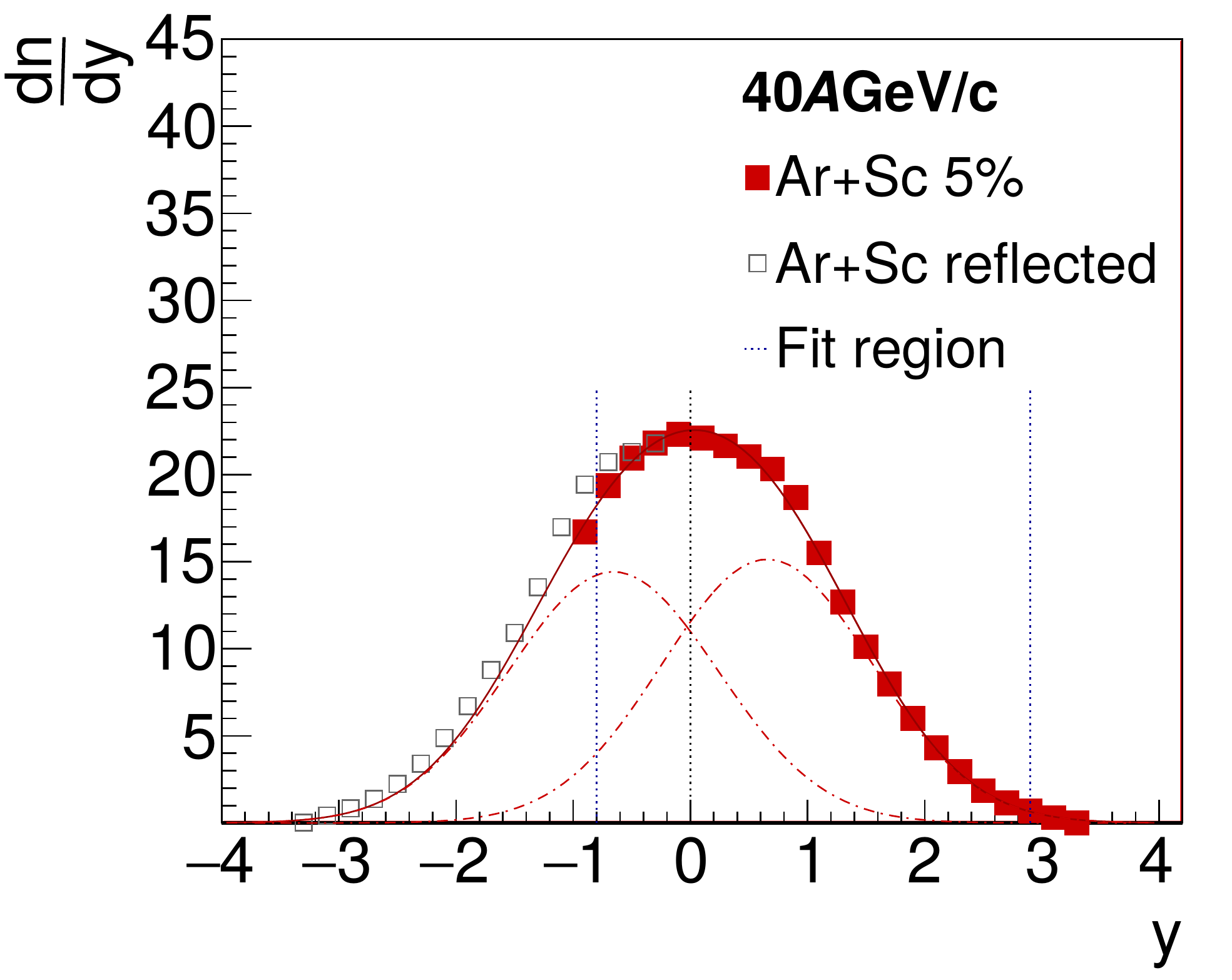}
\includegraphics[width=\twidth]{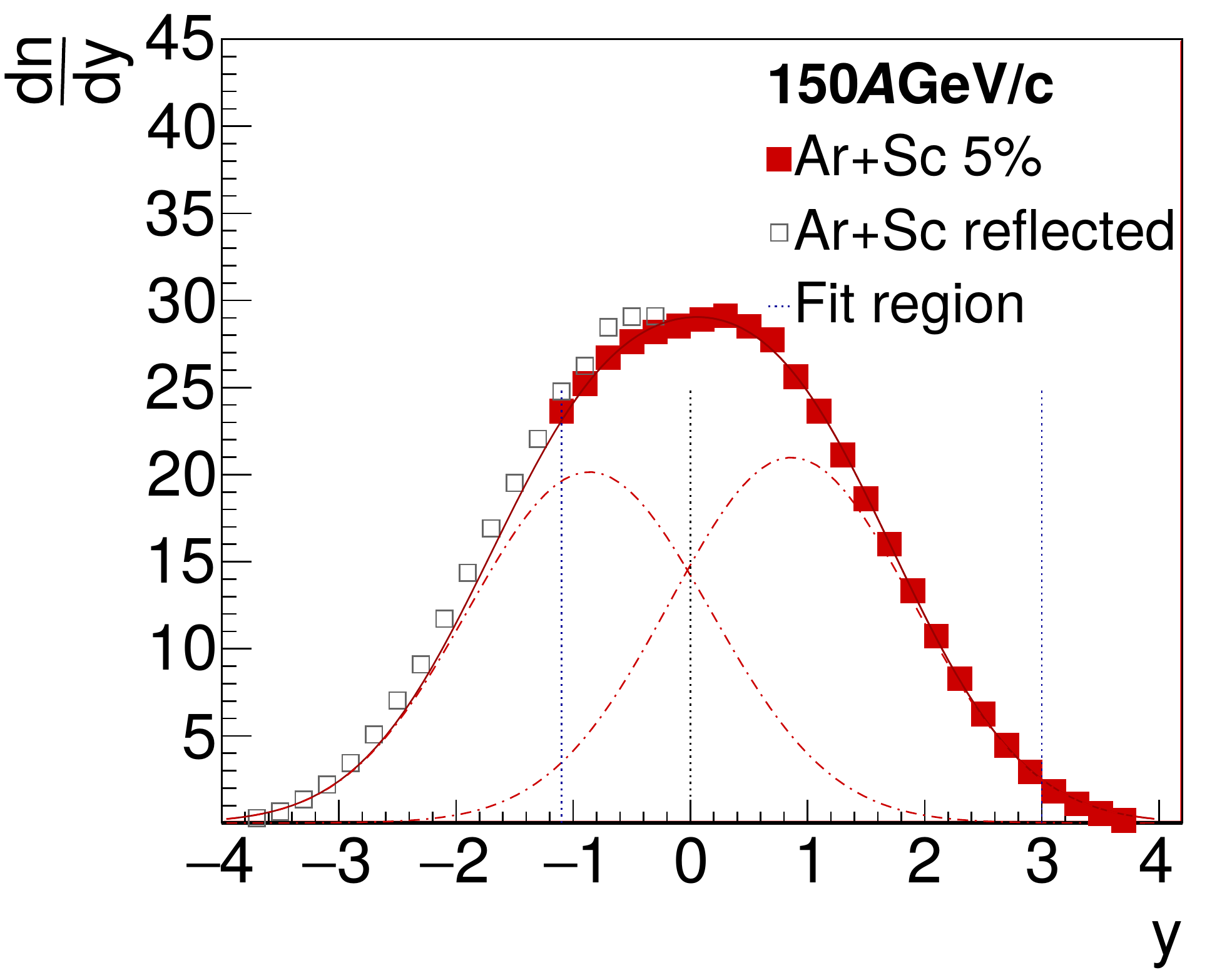}
\caption{\textit{Rapidity spectra  of negatively charged pions in central Ar+Sc collisions.}}
\label{fig:rap_gaussians}
\end{figure}\\
\indent Resulting rapidity spectra (see Fig.~\ref{fig:rap_gaussians}) are asymmetric with respect to mid-rapidity. Thus they were fitted with two Gaussian functions, placed symmetrically around mid-rapidity, with different amplitudes:
\begin{equation}
f_1 = \frac{A_0 A_{rel}}{\sigma_0 \sqrt{2\pi}} \exp \left( -\frac{(y-y_0)^2}{2\sigma_0^2} \right)
,~~~
f_2 = \frac{A_0}{\sigma_0\sqrt{2\pi}} \exp \left( -\frac{(y+y_0)^2}{2\sigma_0^2} \right)
\label{two_gauss}
\end{equation}
$$
f_{fit} = f_1 + f_2~,\hspace{0.2\textwidth} \sigma = \sqrt{\sigma_0^2 + y_0^2}
$$
\begin{figure}[h]
\includegraphics[width=0.4\textwidth]{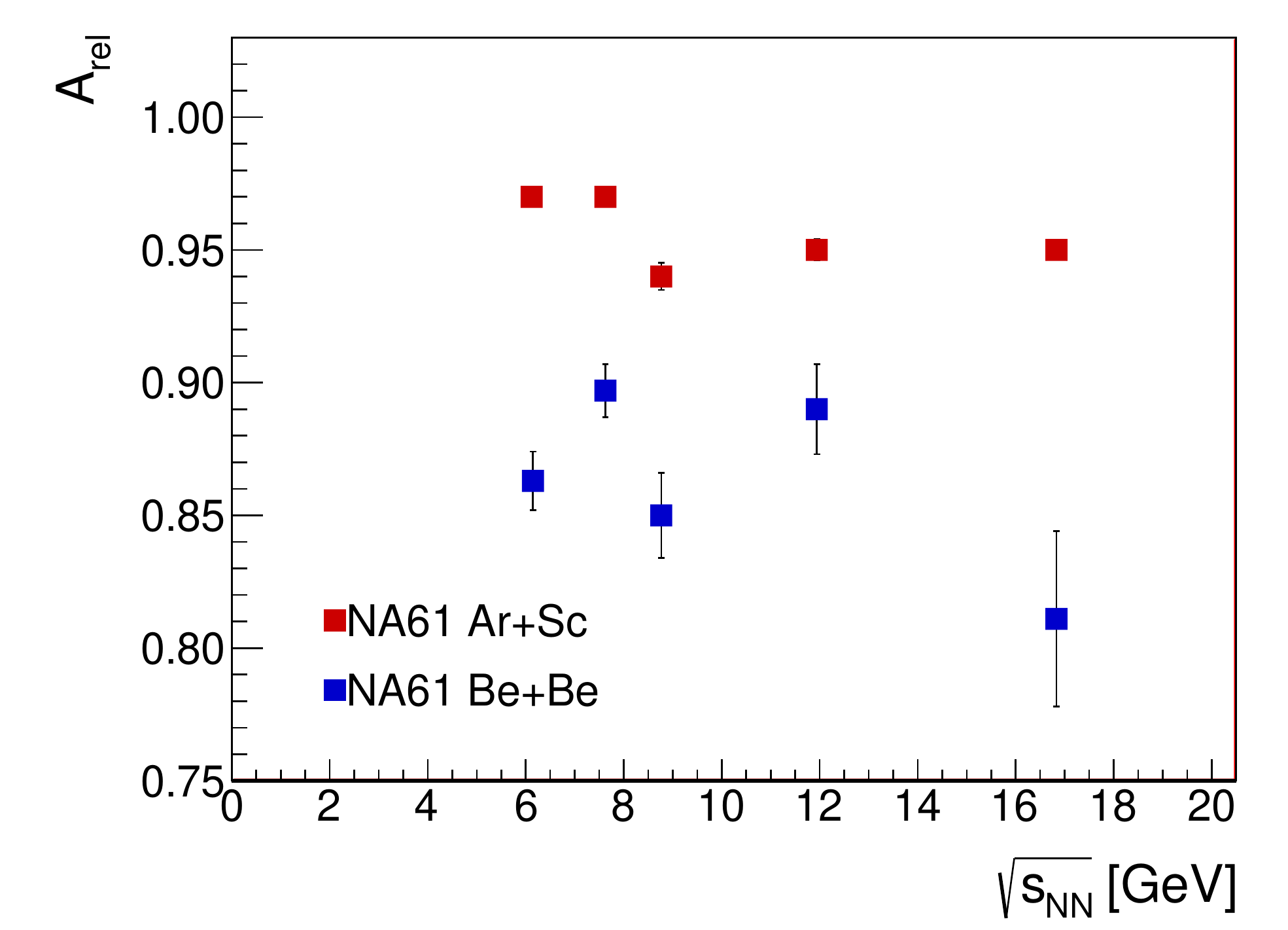} \includegraphics[width=0.4\textwidth]{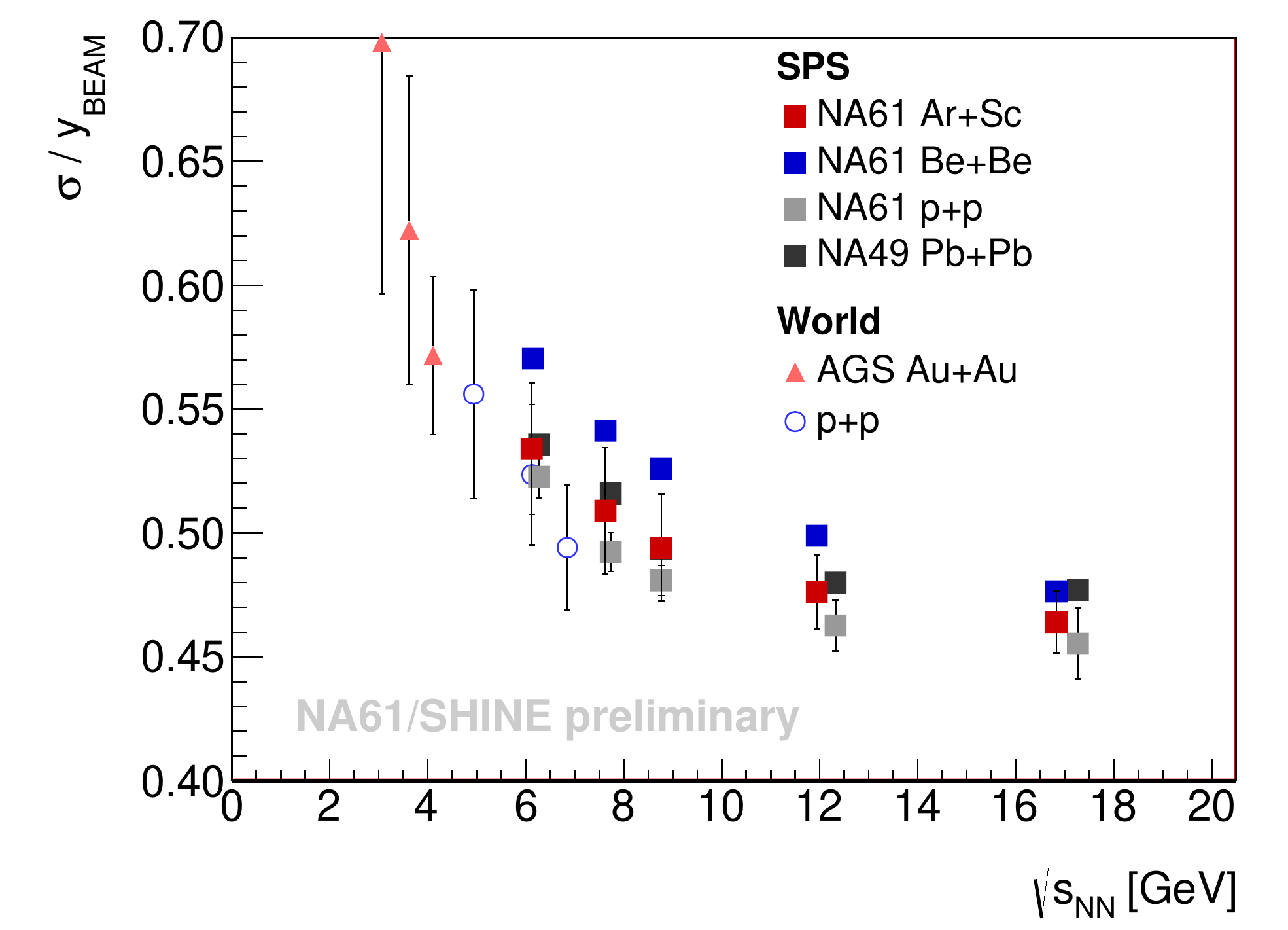}
\caption{\textit{Results of fits of the rapidity distributions with two Gaussians Eq.~\ref{two_gauss}: relative amplitude (left) and width (right). Measurements for Ar+Sc are shown and compared to other systems.}}
\label{fig:fit_params}
\end{figure}\\
\indent Two opposiing effects influence the asymmetry of the spectra: the asymmetry of the system ($^{40}$Ar projectile on $^{45}$Sc target) and the centrality selection based on projectile spectators. The effect of the projectile-spectator asymmetry ($A_{rel}>1$) is overcome by the larger effect introduced by centrality selection ($A_{rel}<1$) (see Fig.~\ref{fig:fit_params}~left). The width $\sigma$ of the rapidity distributions shows monotonic behavior with increasing energy (see Fig~\ref{fig:fit_params}~right).\\
\indent However, non-monotonic behavior is observed for the dependence of $\sigma$ on the system size (Fig.~\ref{fig:fit_params}~right):
\begin{center}$
\frac{\sigma_y(\textrm{p+p)}}{y_{\textrm{beam}}} < \frac{\sigma_y(\textrm{Ar+Sc)}}{y_{\textrm{beam}}} \approx \frac{\sigma_y(\textrm{Pb+Pb)}}{y_{\textrm{beam}}} < \frac{\sigma_y(\textrm{Be+Be)}}{y_{\textrm{beam}}}
$\end{center}
Results for p+p collisions, however, are not yet corrected for isospin effects.\\
\indent In order to investigate the system size dependence of the shape of the $\pi^-$ rapidity spectra $\frac{dn}{dy}$  the results from central nucleus-nucleus interactions are divided by those from inelastic p+p interactions. The spectra were normalized to the same integral in $y\in(0.0,0.5)$ before division.
\begin{figure}[h]
\begin{center}
\includegraphics[width=\twidth]{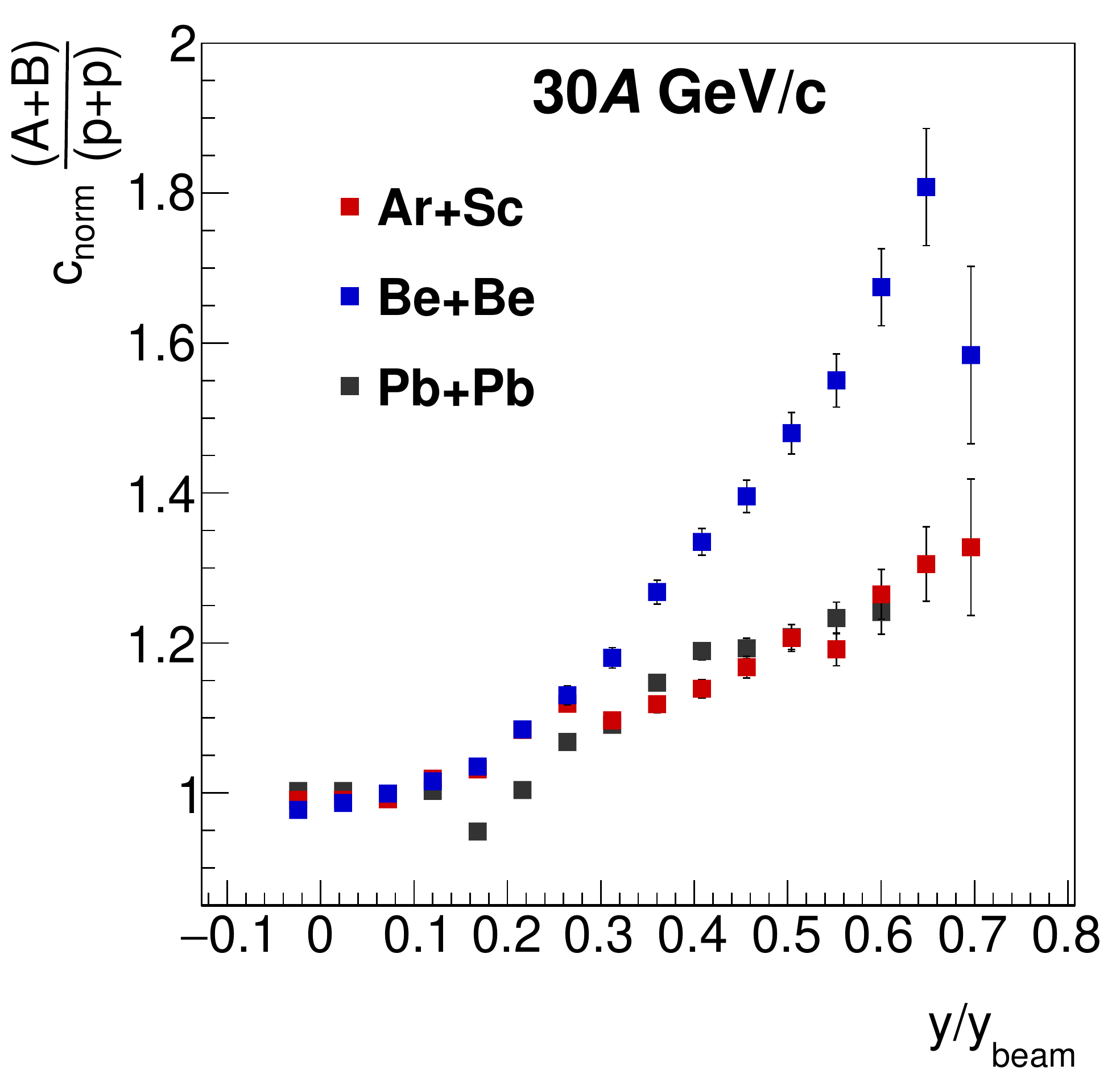} \includegraphics[width=\twidth]{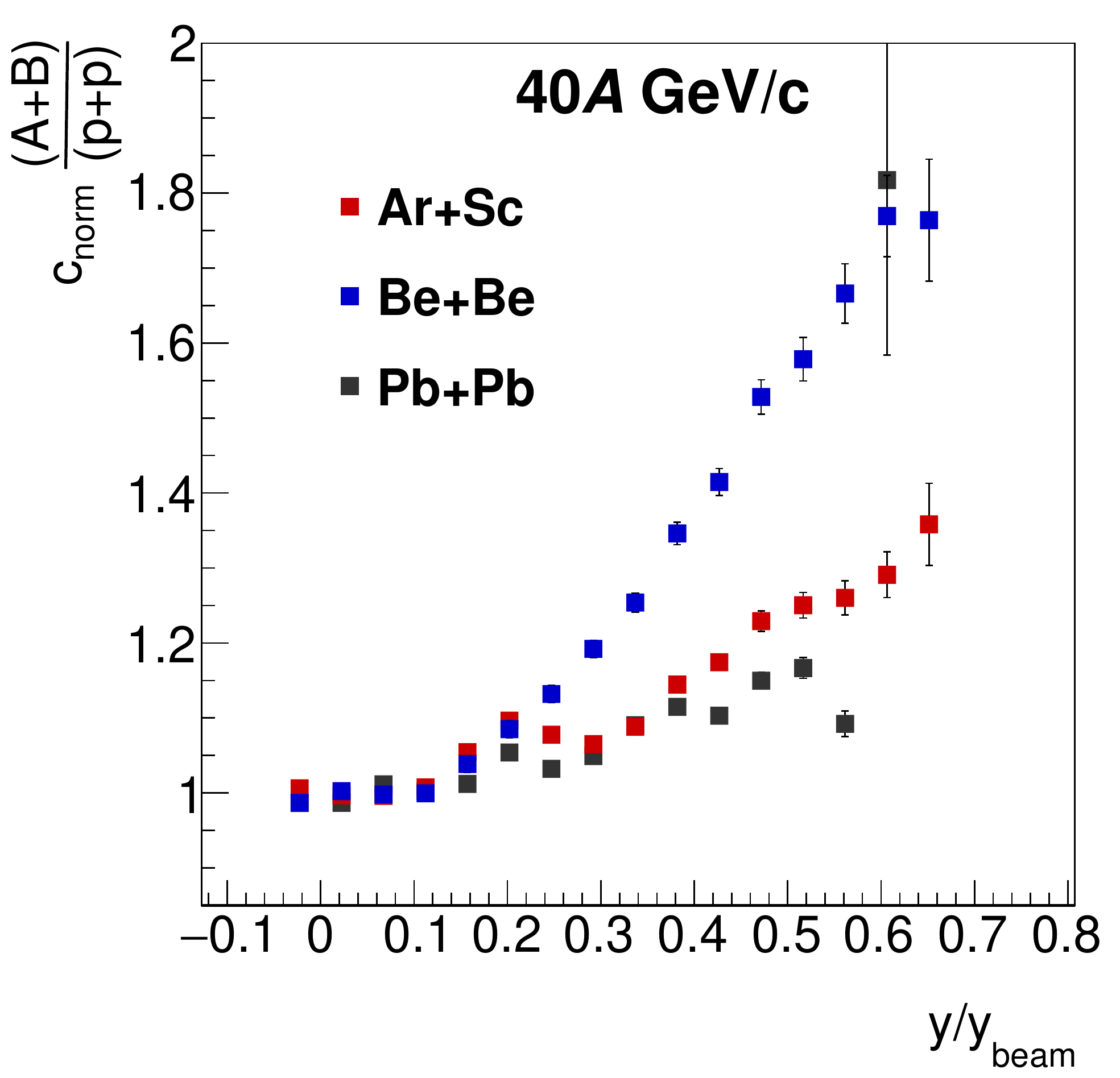}
\includegraphics[width=\twidth]{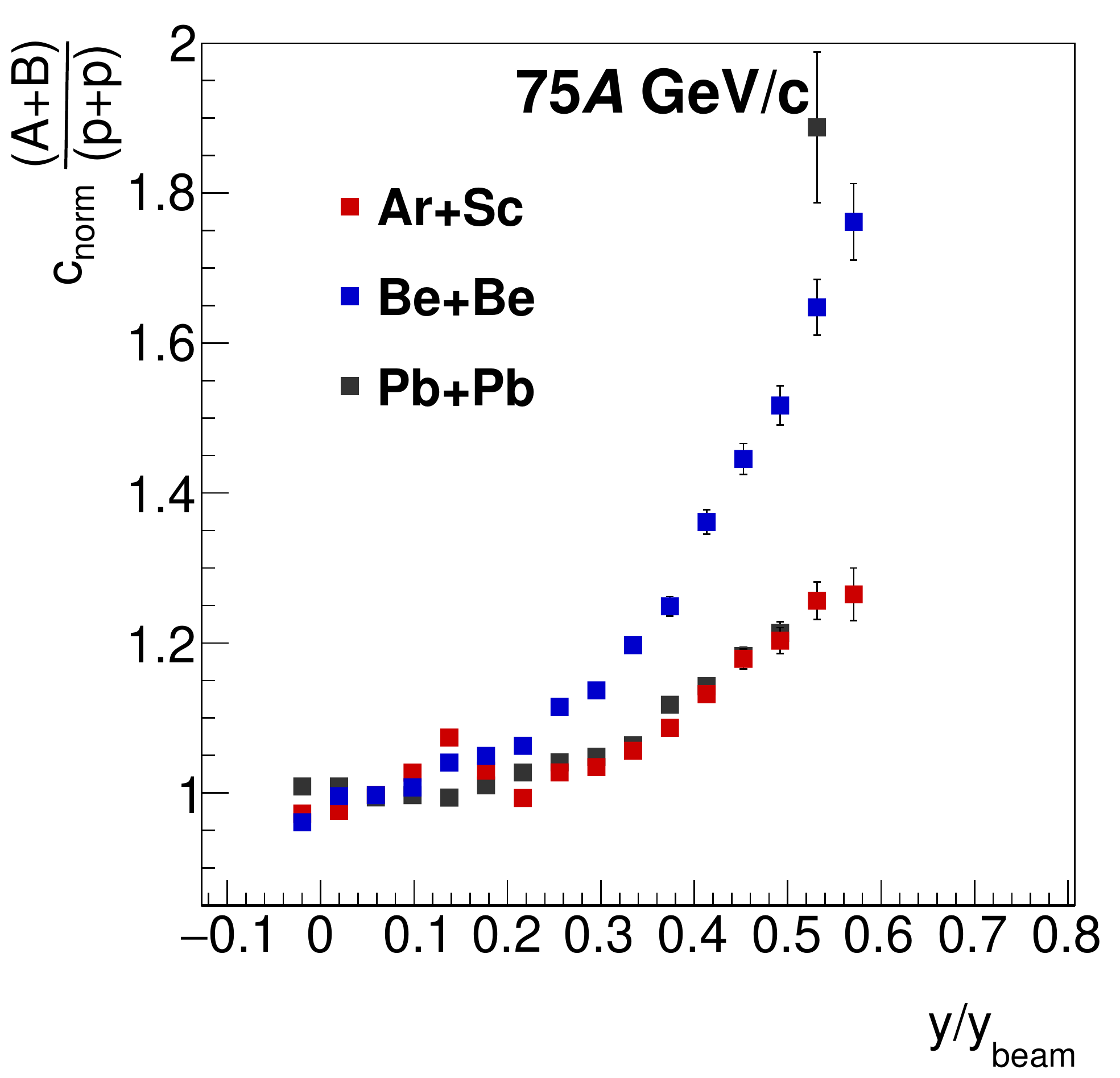}
\end{center}
\caption{\textit{Ratio of $\frac{dn}{dy}$ spectra of $\pi^-$ produced in central nucleus-nucleus collisions to that from inelastic p+p reactions.}}
\label{fig:y_ratios}
\end{figure}\\
From the results plotted in Fig~\ref{fig:y_ratios} one observes that the spectrum shape of $\pi^-$ produced in Ar+Sc interactions is almost identical to the one from Pb+Pb, while data for Be+Be drift away significantly.
\section{Transverse mass distribution} 
\begin{figure}
\begin{center} \includegraphics[width=\twidth]{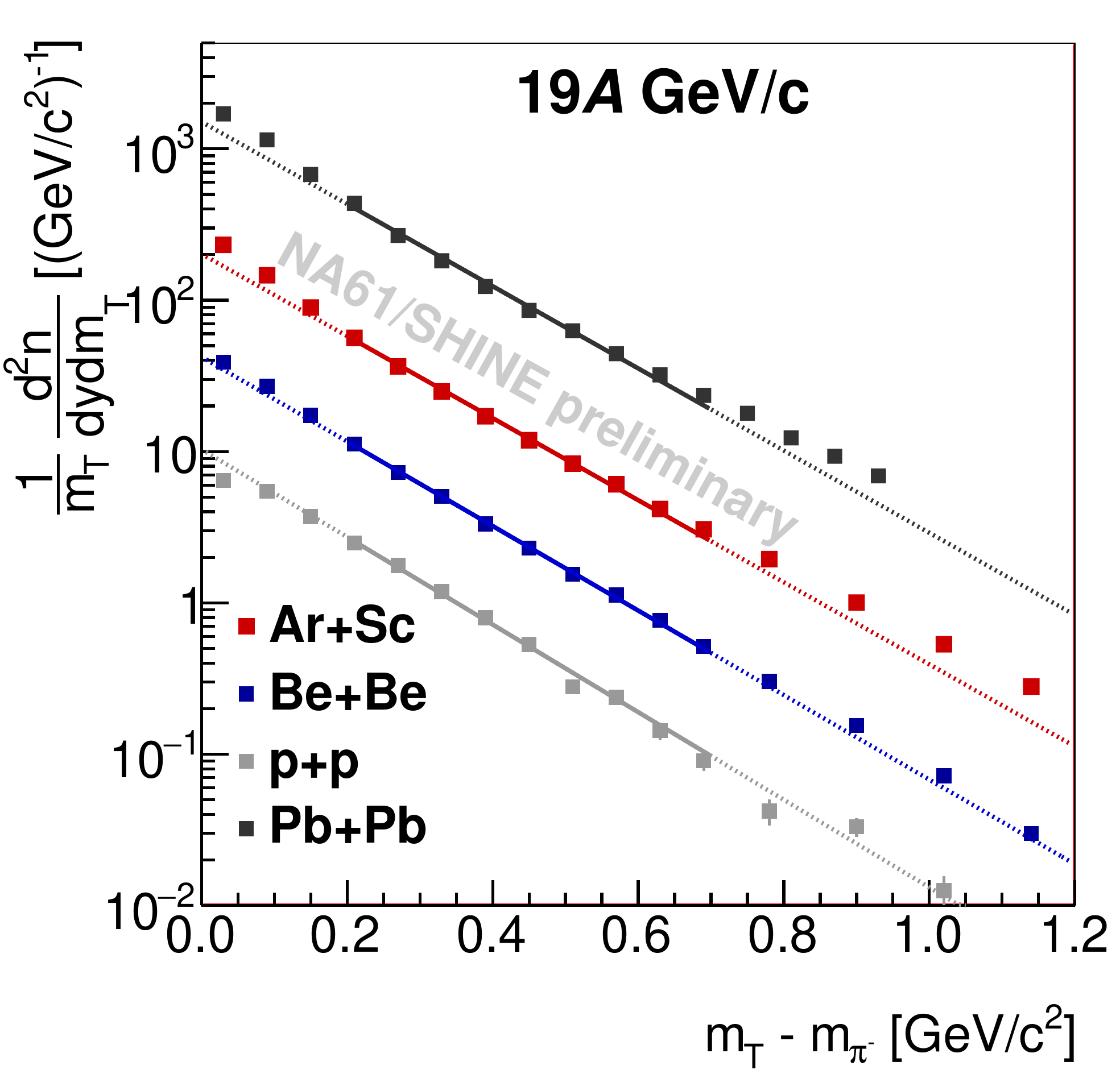} \includegraphics[width=\twidth]{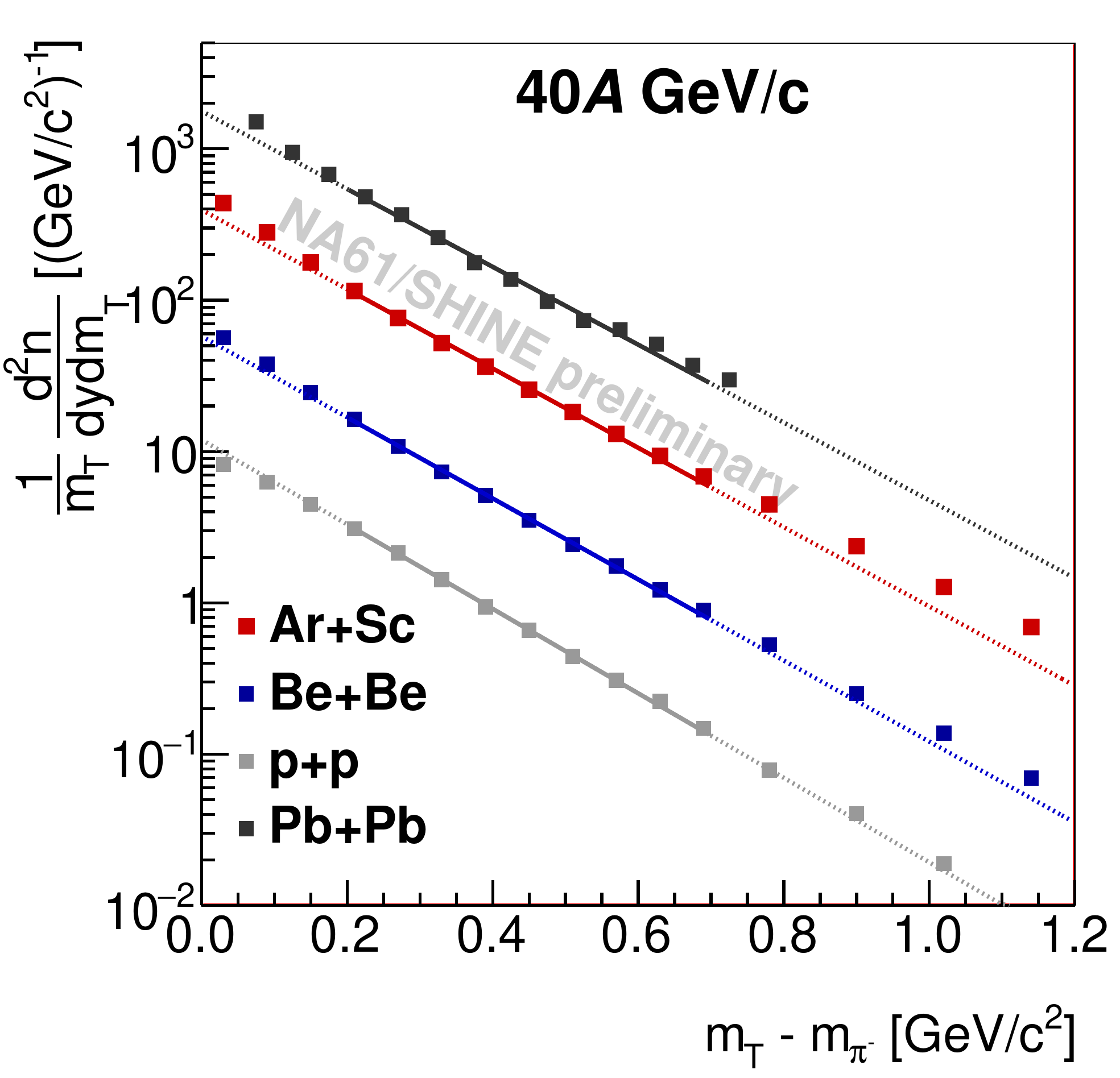} \includegraphics[width=\twidth]{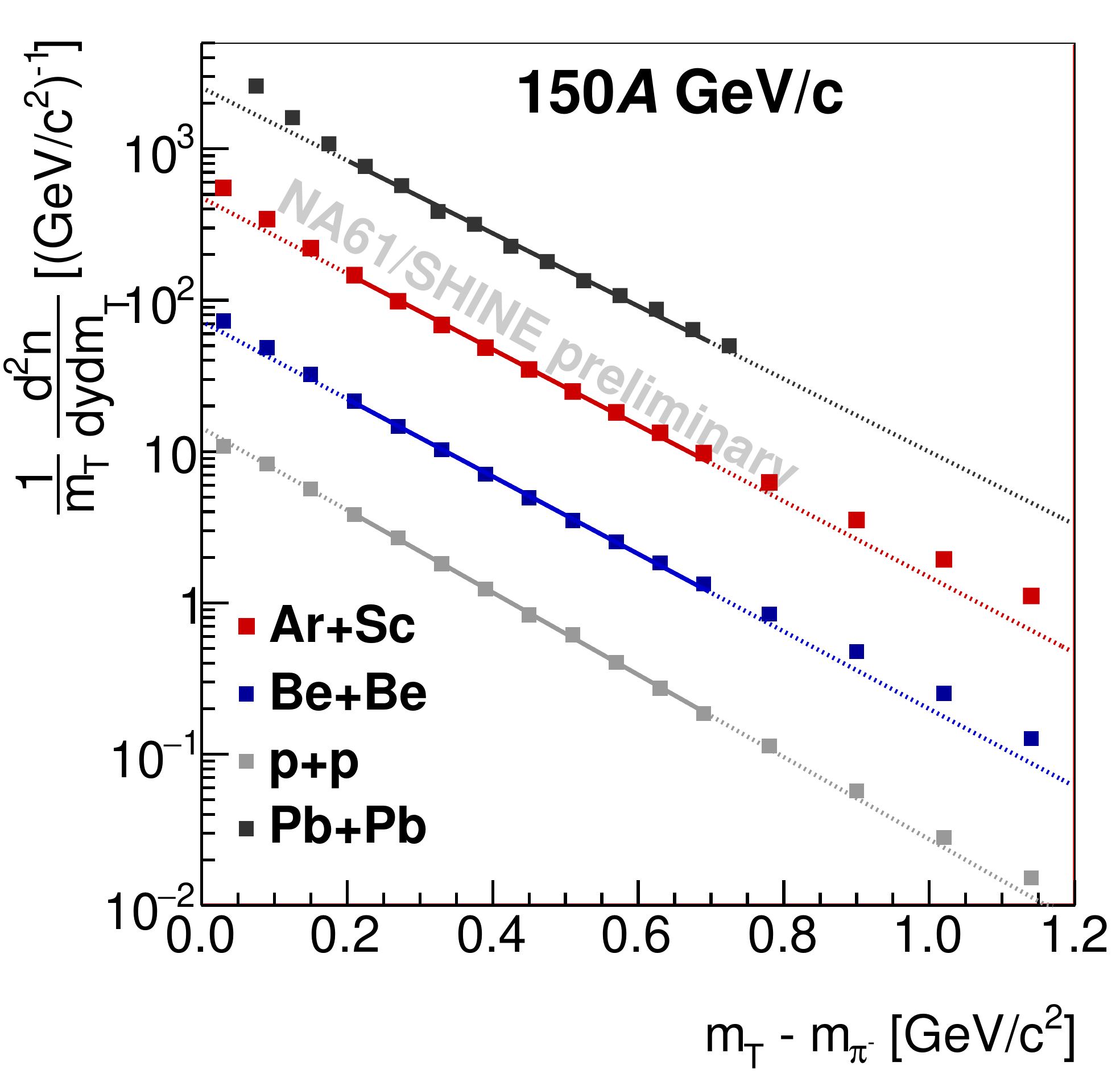}
\end{center}
\caption{\textit{$m_T$ spectra of $\pi^-$ produced in Ar+Sc, as well as other central nucleus-nucleus interactions. Lines depict results of exponential fits in $(m_T-m_{\pi^-})\in(0.2,0.7)$.}}
\label{fig:mt_spectra}
\end{figure}

\begin{figure}
\begin{center}
\includegraphics[width=\twidth]{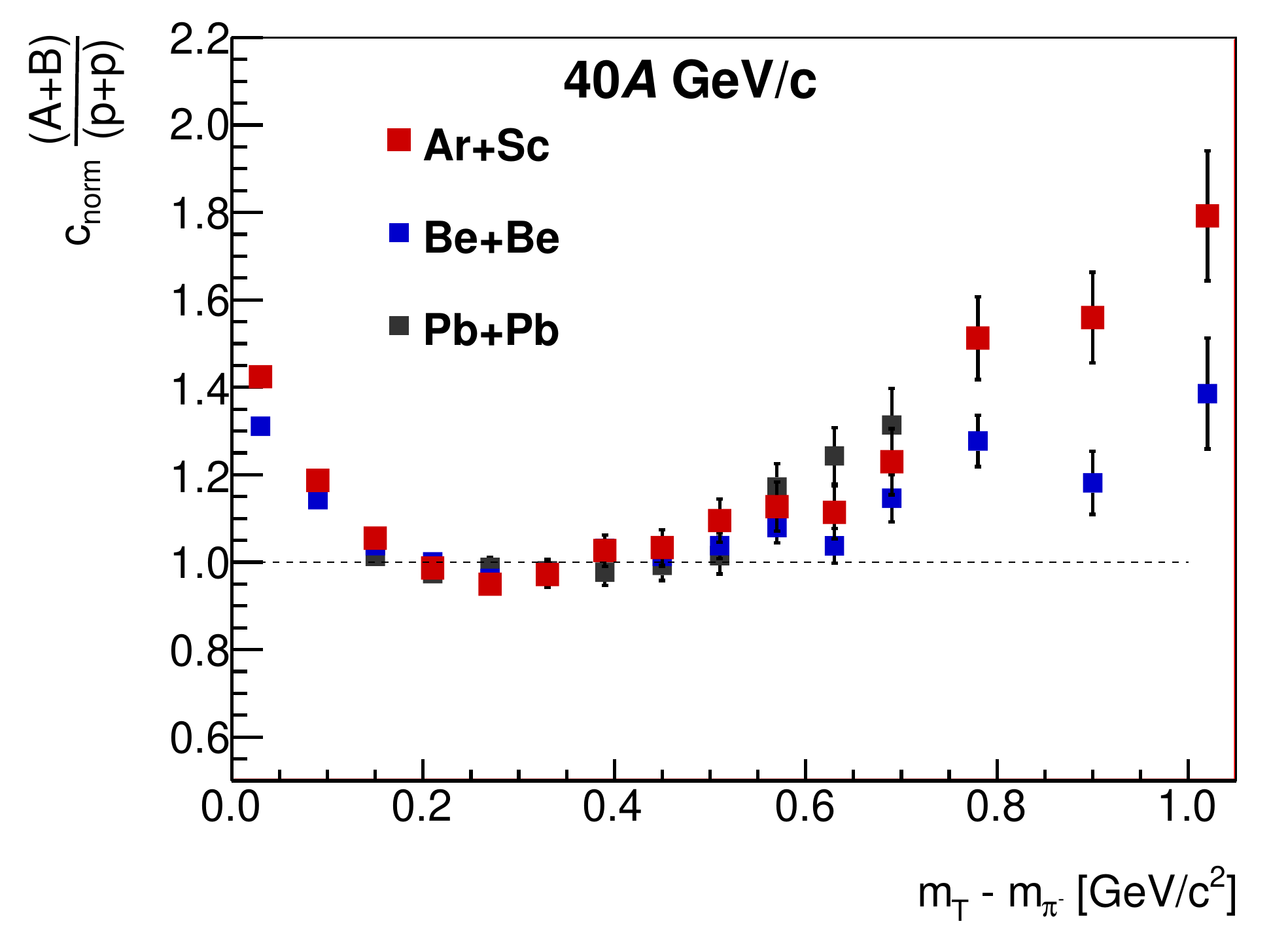} \includegraphics[width=\twidth]{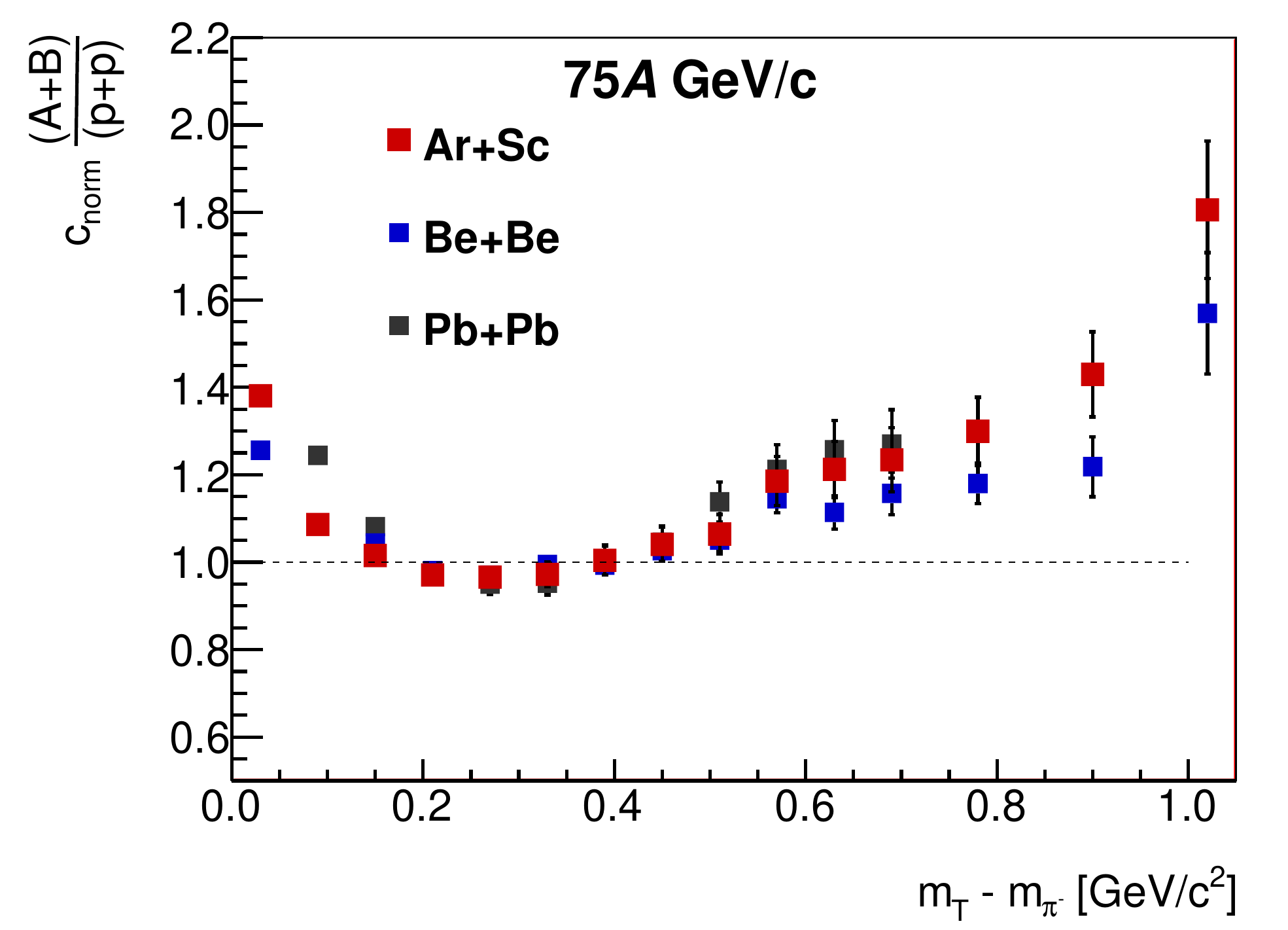} \includegraphics[width=\twidth]{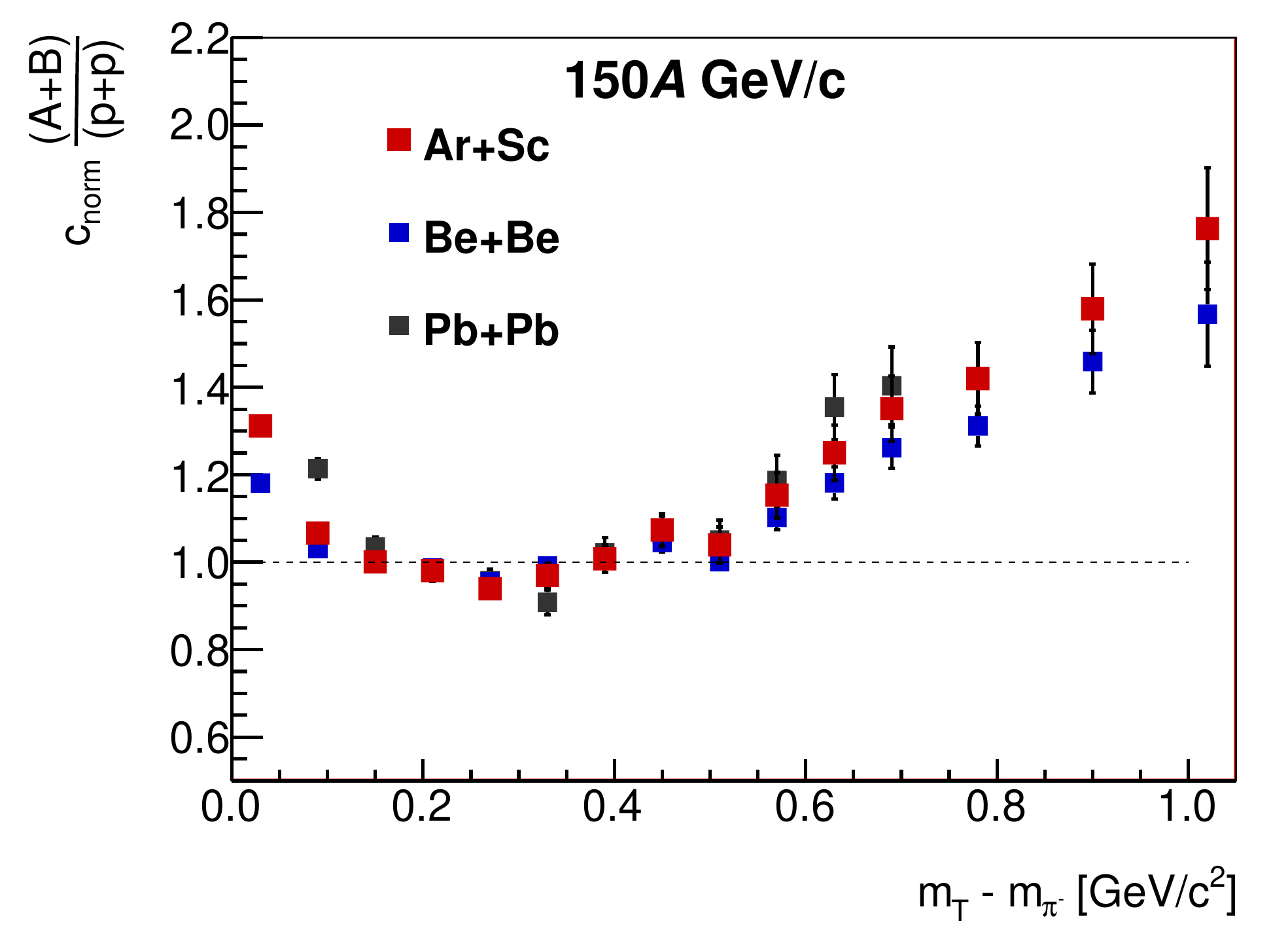}
\end{center}
\caption{\textit{$m_T$ spectra of $\pi^-$ in nucleus-nucleus collisions divided by the p+p data. The spectra were normalized to the integral in $(m_T-m_{\pi^-})\in(0.2,0.7)$ before dividing.}}
\label{fig:mt_ratios}
\end{figure}

Transverse mass spectra of $\pi^-$ for Ar+Sc interactions, shown in Fig.\ref{fig:mt_spectra}, show deviations from the exponential behaviour seen in p+p collisons. These indicate effects of radial flow as seen for Pb+Pb and Be+Be interactions.\\
\indent Another observation on transverse mass spectra is that the deviation from the p+p data at high $m_T$ is larger for heavier nuclei, while there is no observable dependence on collision energy (see Fig.~\ref{fig:mt_ratios}).
\\
\begin{center} \textbf{References} \end{center}
\vspace{-1cm}
\renewcommand\refname{}
\vspace{-1cm}
\bibliographystyle{unsrt}

\end{document}